\documentstyle[subeqnarray]{article}

\def\al{\alpha}

\def\be{\begin{equation}}
\def\ee{\end{equation}}
\def\bea{\begin{eqnarray}}
\def\eea{\end{eqnarray}}
\def\la{\label}

\def\bsea{\begin{subeqnarray}}
\def\esea{\end{subeqnarray}}
\def\u{\underline}

\def\Ó{"}
\def\Õ{' }
\begin{document}
\title{An exactly solvable toy model that mimics the mode coupling theory of supercooled liquid and glass transition}

\author{Kyozi Kawasaki  \\
       Department of Natural Science and Mathematics, \\
       Chubu University,  Kasugai, Aichi 487-8501, Japan\\  \and
Bongsoo Kim \\
       Department of Physics, Changwon National University \\
Changwon, 641-773, Korea  }
\maketitle
\begin{abstract}
A toy model is proposed which incorporates the reversible mode coupling mechanism
responsible for ergodic-nonergodic transition with trivial Hamiltonian in the mode coupling theory (MCT) of structural glass transition. 
The model can be analyzed without relying on uncontrolled approximations inevitable in the current MCT. The strength of hopping processes can be easily tuned and the ideal glass transition is reproduced  only in a certain range of the strength. On the basis of the analyses of our model we  discuss about a sharp ergodic-nonergodic transition and its smearing out by "hopping".\\
\medskip
\noindent
{\it PACS:05.40.+j;51.10.+y;61.20.-p;64.70Pf}
\end{abstract}
\vspace{0.5cm}

The mode coupling theory(MCT) as applied to supercooled liquid and glass transition is the only existing first principle theory for this last stronghold of condensed matter physics with surprising success in describing the initial stages of freezing\cite{gs}. However, as one expects, there are lots of formidable issues in applying the theory originally designed for critical phenomena dealing with thousands of \AA ngstroms to the short scale  problems of at most tens of \AA ngstroms. In the original derivation of the idealized MCT self-consistent equation for the density-density correlator, the factorization approximation of replacing the four-body time correlation function by the product of two-body time correlation functions was introduced, which is totally uncontrolled and whose region of validity is unknown. Also nature of the non-ergodic state after the freezing transition is far from clear. The difficulty is compounded if one goes over to  the so-called extended MCT where the rapidly varying momentum variable is introduced in the MCT scheme, which was necessary to partially cover the thermally activated processes important after freezing transition has taken place. However, such rapidly varying momentum variable can never be sensibly treated by MCT\cite{lo97}. Furthermore, physical picture of the hopping processes is still lacking.

In the recent years, the ideas and the methodology developed in the spin glass community are brought in to deal with structural glass problems, which was  pioneered by Kirkpatrick, Wolynes and Thirumalai\cite{kwt}. Here mean field type toy models like $p$-spin models are being analyzed producing deep insights. However, in all these spin glass models the glass transitions are driven by non-trivial Hamiltonians (or free energy functionals)\cite{sg}.

 On the other hand, in the above-mentioned MCT for structural glass, the transition is driven by the {\em reversible mode coupling mechanism}, and does not require a nontrivial Hamiltonian. Considering these circumstances we feel it important to develop a mean field type toy model that mimics this reversible mode coupling mechanism with trivial Hamiltonian and that can be easily analyzed. Proposition of such a toy model  is the purpose of this communication. The model can be exactly solved and yield the self-consistent equation for the relevant correlator of the type familiar in the mode coupling theories of supercooled liquid and glass transition, where the strength  of hopping processes  can be readily tuned.

Our model contains the $N$-component density variable $a_i$ with $ i=1,2,3, \cdots N$ and the $M$-component velocity field variable  $b_{\al}$ with $ \al=1,2,3, \cdots M$. In the end $N$ and $M$ will tend to $\infty$ keeping the parameter $\delta^* \equiv M/N$ finite. The model equations of motion are expressed by the following Langevin-type equations for nonlinear random coupled oscillators with damping and noise:
\bsea
\dot{a}_i&=& K_{i\alpha}b_{\alpha}+\frac{\omega}{\sqrt{N}}
J_{ij\alpha}  a_j b_{\alpha}  \\ \dot{b}_{\alpha} &=&
-\gamma b_{\alpha}-\omega^2 K_{j\alpha
}a_j-\frac{\omega}{\sqrt{N}}J_{ij\alpha} (\omega^2a_ia_j-
T\delta_{ij})+f_{\alpha}   \\
 <f_{\alpha}(t)f_{\beta}(t')> &=&
2\gamma T\delta_{\alpha \beta} \delta (t-t')
\la{eqn:s6}
\esea
where  the upper dot is the time derivative and the angular bracket is the average over the Gaussian white thermal noise denoted by $f_{\alpha}$. Here and after we use the summation convention over repeated indices.   $\omega,\gamma$ and $T$ are the positive constant parameters of the model and $K_{i \al}$ are the coefficients satisfying the conditions $ K_{i\al}K_{i \beta}=\delta_{\al \beta}$ ( but {\it not} the conditions  $ K_{i\al}K_{j \al}=\delta_{ij}$ for $M<N$). $J_{ij\alpha}$ are the mode coupling coefficients which are chosen to be "quenched" Gaussian random variables with the following properties:
\bsea
\overline{J_{ij\alpha}}^J &=&0,  \\
\overline{J_{ij\alpha}J_{kl\beta}}^J &=& \frac{g^2}{N}
\biggl[(\delta_{ik}\delta_{jl}+\delta_{il}\delta_{jk})\delta_{\alpha
\beta}+K_{i\beta}(K_{k\al}\delta_{jl}+K_{l\al}\delta_{jk})
\nonumber \\ &+&
K_{j\beta}(K_{k\al}\delta_{il}+K_{l\al}\delta_{ik})\biggr]
\la{eqn:s5}
\esea
where $\overline{\cdots }^J$ denotes average over the $J$'s.

(\ref{eqn:s6}a) is analogous to the equation of continuity of fluid and (\ref{eqn:s6}b) is like the equation of motion where the rhs is like
the force acting on a fluid element. The coefficients $K_{i\al}$ govern linearized reversible dynamics of the oscillators with the dynamical matrix $\bf \Omega$ given by $\Omega_{ij} \equiv \omega^2K_{i\al}K_{j\al}$. In constructing this model we were motivated by \cite{rk59,bckm96} where random coupling models involving infinite component order parameter have been shown to be analyzed exactly by mean field type concepts.

One can derive from the Langevin equations (\ref{eqn:s6}) the corresponding Fokker-Planck equation for the probability distribution function $D_t(\{a\},\{b\})$ for our variable set denoted as  $\{a\},\{b\}$, whose stationary (equilibrium) solution is given by
\be
 { D}_e(\{a\},\{b\}) = cst. e^{-\sum_{j=1}^N\frac{\omega^2}{2T}a_j^2-
\sum_{\alpha=1}^M\frac{1}{2T}b_{\alpha}^2}
\la{eqn:s4}
\ee
where $cst.$ is the normalization factor.

For the subsequent analysis it is most convenient to adapt the generating functional method 
of \cite{bckm96} to our case, which produces no new problem and its details can be omitted. 
Then below  we  obtain  for this toy model the correlation functions defined by
\bsea
 C_a(t,t') &\equiv&  \frac{1}{N} <a_j(t)a_j(t')>, \quad
 C_{ab}(t,t') \equiv \frac{1}{M}K_{j\al}<a_j(t)b_{\al}(t')>,
  \\
 C_{ba}(t,t') &\equiv& \frac{1}{M}K_{j \al}<b_{\al}(t)a_j(t')>,
 \quad
  C_b(t,t') \equiv \frac{1}{M} <b_{\alpha}(t)b_{\alpha}(t')>\\
  C_a^K(t,t') &\equiv& \frac{1}{M}<a_{\al}^K(t)a_{\al}^K (t')>,\quad
  a_{\al}^K \equiv K_{j\al}a_j
\la{eqn:s14}
\esea
The last one is needed to close the self-consistent set of equations for correlators 
when $M<N$\cite{nn}.

 We can write down the set of
effective linear Langevin equations resulting from the generating functional method which are valid in the limit of $M,N
\rightarrow \infty$:
\bsea
\dot{a}_i(t)&=& K_{i\alpha} b_{\al}(t)- \Sigma_{aa} \bigotimes a_i(t)
-K_{i\al}  \Sigma_{ab}\bigotimes b_{\al}(t)+f^a_i(t)  \\
\dot{b}_{\al}(t)&=&-\gamma b_{\al}(t)-\omega^2
 K_{i\alpha}a_i(t) -K_{i\al } \Sigma_{ba}\bigotimes a_i(t) -
 \Sigma_{bb} \bigotimes b_{\alpha}(t)+ f^b_{\alpha}(t)
 \la{eqn:s32}
\esea
where $f^a$ and $f^b$ are the effective thermal noises whose correlations are related to the $\Sigma$'s in a usual way, and
\be
\Sigma \bigotimes a(t) \equiv \int_{-\infty}^t dt' \,
\Sigma(t,t')a(t') \quad \mbox{etc.}
\ee
Here the kernels $\Sigma$'s are given by
\bea
\Sigma_{aa}(t,t') &\equiv&
\frac{g^2\omega^4}{T}\delta^* \bigl(C_a(t,t')C_b(t,t')+\delta^* C_{ab}(t,t')C_{ba}(t,t')\bigr),
\quad \Sigma_{ab}(t,t') \equiv -2
\frac{g^2\omega^4}{T}\delta^* C_a(t,t')C_{ba}(t,t') \nonumber \\
\Sigma_{ba}(t,t') &\equiv& -2
\frac{g^2\omega^6}{T}\delta^* C_a(t,t')C_{ab}(t,t'), \quad
\Sigma_{bb}(t,t')\equiv
\frac{2g^2\omega^6}{T}C_{a}(t,t')^2.
\la{eqn:s31}
\eea

Using this effective Langevin equation we can derive the equation for correlation functions (\ref{eqn:s14}) which depend only on $t-t'$, 
or their Laplace transforms the $C^L(z)$'s (the $\Sigma^L(z)$'s are the similarly defined Laplace transforms of $\Sigma(t,t')$'s):
\bsea
 zC^L_a(z)&=& \frac{T}{\omega^2} + 
 (1-\Sigma^L_{ab}(z))\delta^* C^L_{ba}(z)- \Sigma^L_{aa}(z)C^L_a(z)  \\
 zC^L_{ba}(z)&=&-(\gamma + \Sigma^L_{bb}(z)) C^L_{ba}(z) -( \omega^2 +\Sigma^L_{ba}(z)) 
C^{KL}_a(z)  \\
 zC^L_{ab}(z)&=& (1-\Sigma^L_{ab}(z)) C^L_b(z)- \Sigma^L_{aa}(z) C^L_{ab}(z) \\
 zC^L_b(z)&=&T - (\omega^2 +\Sigma^L_{ba}(z))C^L_{ab}(z)
-(\gamma + \Sigma^L_{bb}(z)) C^L_b(z)\\
 zC_a^{KL}(z)&=&\frac{T}{\omega^2}+(1-\Sigma^L_{ab}(z))
C^L_{ba}(z)-\Sigma_{aa}^L(z)C_a^{KL}(z)
\la{eqn:s36}
\esea

Since the "self-energies", the $\Sigma$'s, are expressed in terms of the correlators 
by (\ref{eqn:s31}), the equations (\ref{eqn:s36})  
constitute the self-consistent scheme for the 5 correlators (\ref{eqn:s14}). 
These have to be solved numerically under the appropriate initial conditions 
for the $C(t',t')$'s.
Eliminating $C^L_{ba}(z)$ and  $C^{KL}_{a}(z)$ from
 (\ref{eqn:s36}a , b and e) we find the following  equation for  $C^L_{a}(z)$:
\be
C^L_a(z)=\frac{T}{\omega^2}\frac{1}{z+\Sigma^L_{aa}(z)}
 \biggl[1-\omega^2\delta^* \frac{(1-\Sigma^L_{ab}(z))^2}{(z+\Sigma^L_{aa}(z))
(z+\gamma+\Sigma_{bb}(z))+ \omega^2(1-\Sigma^L_{ab}(z))^2} \biggr]
\la{eqn:s38}
\ee
where we  have used the symmetry relation
$\Sigma^L_{ba}(z)=-\omega^2\Sigma^L_{ab}(z)$,  and other correlators entering the $\Sigma$'s 
have been expressed similarly. 

For $\delta^*=0$ the $\{a\}$ variables are time-independent since there is no velocity variable $\{b\}$ that drives dynamics of $\{a\}$.
 Therefore the system is trivially nonergodic: $C_a(t,t')=C_a(t',t')=T/\omega^2$. 
For $\delta^*=1$ (\ref{eqn:s38}) reduces to the equation derived in \cite{sdd93} and the system is ergodic due to the presence of strong hopping process.
For intermediate values of $\delta^*$ we may expect ergodic-to-nonergodic transition
 as we have numerically seen for $\delta^*=0.3$ (see Fig.1). 
For small values of $\delta^*$ we have the following result correct up to first order
 in  $\delta^*$:
\be
C^{L}_a(z)= \frac{T}{\omega^2}\frac{1}{z}\biggl[ 1 -\delta^*
\frac{1+g^2T}{1+2g^2T}\biggr]
\la{eqn:s39}
\ee
The phase diagram in the variables $\delta^*$ and $T$  found numerically is displayed in Fig. 2. Equating (\ref{eqn:s39}) to zero we find the transition at $T_c=g^{-2}(1-\delta^*)/(\delta^*-2)<0$, the unphysical value. Hence we  suppose that $T_c$ should diverge at some nonvanishing positive value of $\delta^*$.

It should be emphasized that our theory is {\em exact} in
the limit of large $M$ and $N$. By varying  the parameter $\delta^*$
which is the ratio of the numbers of components of $\{a\}$ and $\{b\}$,
we can control the strength of the hopping processes represented by the  $\Sigma_{aa}$
and  $\Sigma_{ab}$ terms.  Reducing $\delta^*$ drives the system into
glassy region.

Our na\"ive expectation when we started this work which was for $\delta^*=1$
({\it i.e.} $M=N$ and $K_{i\al}=\delta_{i\al}$) was the following. 
We are constructing an $N$-component model for which 
 mean field mode-coupling approximation is exact as $N \rightarrow \infty$.
 A mean field type approximation often gives a sharp transition and G\"otze' s MCT without 
"hopping" can be regarded as a kind of mean field theory. 
Hence it is natural to expect that this toy model designed to reproduce 
the G\"otze type mean field theory rigorously  has such a sharp transition.
 But unexpectedly our toy model did not show a sharp transition.
 In order to produce a transition it was necessary to generalize 
the original $N$-component model to allow $M<N$.
 Thus, understanding this so-called ideal glass transition without relying on
 uncontrolled approximation appears to be more subtle than expected.
 Another related issue is concerned with the nature of the "hopping" process which
 can occur  even with trivial Hamiltonian.
 This means  that the "hopping"  found in our model has nothing to do with
 thermally activated energy barrier crossing since the trivial Hamiltonian does not
 possess such a barrier.  The same comment applies to  \cite{sdd93}.
 Thus we are faced with the problem of understanding the Arrhenius type hopping rate
 found recently\cite{gv00}.

One may interpret that since the reduction of $\delta^*$ is tantamount to
 restricting accessible region of the phase space,
 we can say that we have an entropic barrier\cite{backg}.
 This interpretation, however, need to be substantiated. In this connection it is illuminating to eliminate the variables $\{b\}$ adiabatically to obtain a closed Fokker-Planck type equation containing only the $\{a\}$ variables. Then we can show that the diffusion marix in this equation is singular\cite{hr89}, {\it i.e.} its determinant vanishes. This gives rise to a possibilty that the equation can have stationary solutions other than the equilibrium Gaussian one\cite{hr89}. Such solutions are precisely the kind of non-ergodic states found numerically for the present original model. This feature will be signifincant beyond the present toy model. We also note the simultaneous appearance of the ergodic to non-ergodic transition and the new variable $  a_{\al}^K \equiv K_{j\al}a_j$, (\ref{eqn:s14}c). Elucidating its implication is an intriguing future task.

 We hope to discuss these aspects further including critical behavior near transition and an extension to aging in future.

K.K. acknowledges  supports of the following research project and funding agencies: the Cooperative Research under the Japan-U.S. Cooperative Science Program,  Scientific Research Fund of Ministry of Education, Science and Culture of Japan, and   Research Institute for Science and Technology of Chubu University.
B.K. was supported by grant No. 1999-2-114-007-3 from the Interdisciplinary Research
Program of the KOSEF.
A part of this work was carried out when both authors spent two summers  at Kyushu University.
We would like to thank hospitality of Prof. S. Kai who provided office spaces and access to the facilities of his laboratory.

\medskip
\u{Figure Captions}\\
Fig.1  Temporal behavior of $C_a(t)\equiv C_a(t,0)$ for $g=\gamma=\omega=1,\, \delta^*=0.3.$. 
The curves are, from left to right at long times, 
for $T=1.0,\,0.5, \, 0.3, \, 0.2,\,0.1,\,0.05, \, 0.01, \, 0.001$.\\
Fig.2  The phase diagram in the $\delta^*-T$ space dividing the ergodic and non-ergodic regions. 

\begin{thebibliography}{99}
\bibitem{nn} For $M=N$ we can impose an additional 
condition $K_{i\al}K_{j\al}=\delta_{ij}$, which results in
 $C_a^K(t,t')=C_a(t,t')$.
\end{thebibliography}
\end{document}